\documentclass {article22}
\usepackage[english]{babel}
\usepackage{graphicx}
\usepackage{makeidx}
\usepackage{amsmath,natbib}

\begin{document}

\begin{frontmatter}

\title{A unique radioisotopic label as a new concept for safeguarding and tagging of long-term stored items and waste}

\author[rvt]{Dina Chernikova\corref{cor1}}
\author[rvt,focal]{K{\aa}re Axell}
\cortext[cor1]{Corresponding author, email: dina@nephy.chalmers.se}
\address[rvt]{Chalmers University of Technology, Department of Applied Physics, Nuclear Engineering, \\Fysikg{\aa}rden 4, SE-412 96 G\"oteborg, Sweden}
\address[focal]{Swedish Radiation Safety Authority, SE-171 16 Stockholm, Sweden}

\begin{abstract}

The present paper discuss a novel method of tagging and labeling of waste casks, copper canisters, spent fuel containers, mercury containers, waste packages and other items. In particular, it is related to the development of new long-term security identification tags/labels that can be applied to articles for carrying information about the content, inventory tracking, prevention of falsification and theft etc. It is suggested to use a unique combination of radioisotopes with different predictable length of life, as a label of the items. The possibility to realize a multidimensional bar code symbology is proposed as an option for a new labeling method. The results of the first tests and evaluations of this are shown and discussed in the paper. The invention is suitable for use in items assigned to long-term (hundreds of years) storing or for final repositories. Alternative field of use includes fresh nuclear fuel handling and shipment of goods.

\end{abstract}


\end{frontmatter}

\section{Introduction}
\emph{"...The desire to uniquely identify valuables is not new to human thought. Written history shows that the ancient Egyptians used seals to identify government documents, while the Babylonians used tags and seals for their trade with the Indian and Chinese civilizations"}, - Christos Makris (DOE Office of Research and Development) \cite{DOE}. Thousands of years has passed since then, technology progresses, and even despite this tags and seals did not loose their actuality and importance. Nowadays security and identification labeling are extensively used in everyday life and industry to track containers and products, in automatic and manual ways. There are several reasons for using identification tags, such as to provide the verification of the items in question, to identify the theft or the misuse of an item and to provide the information about the item without breaking its integrity. Therefore, while choosing a particular type of tag it is necessary to consider a number of important parameters.

    \subsection{Main requirements to the ideal tagging system}
There were a few attempts to systematize criteria for the selection of a specific tag, for example, based on: purpose of tag, type of the container, robustness, reliability, easy of application, effectiveness, interface with other safeguards and security elements, cost etc. \cite{DOE}. Although, in connection with a long-term (hundreds of years) stored item, such as nuclear waste, spent fuel or mercury containers, one can consolidate these requirements in five main points ("intuitive requirements"), i.e. the ideal tag must provide:
\begin{enumerate}
  \item Environmental safety (avoid corrosion effects of e.g. copper canisters).
  \emph{The labeling system should avoid corrosion effects of canisters which can be induced in the long-term run, thus for instance avoiding leakage of spent fuel waste components later on.}
  \item Non-contact reader system (preferably).
  \item Long operation time.
  \emph{The labeling system should have an operating time at least from ten to a few hundred years.}
  \item Large and unique tag memory.
     \emph{The labeling system should enable fully unique identification of the canister content in a manner consistent with permanent records of the storage or repository.}
  \item Security technique against falsification of data, errors/multiple verification.
 \emph{ The labeling system should have high level of security, i.e. low risk of falsification or error.}
\end{enumerate}

Thus, an ideal identification tag meet all the challenges of the international initiative on a holistic Safety, Security and Safeguards ('3S') concept. Therefore, hereafter we will consider the suitability of the currently existing technologies and new approach to these "intuitive requirements".
    \subsection{Overview of existing technologies}
The conventional tagging techniques include etching characters, affixing identification plates, welding, etc. However, when considering an application for long-term storage of waste canisters they have a number of gaps in the factors of environmental safety, security and long operation time. Other disadvantages of the traditional labeling technology are described in \cite{1}.
Modern labeling techniques may partly solve these problems and be useful for meeting the goals of a unique labeling system compatible with the record keeping of the storage or repository. Among the modern labeling techniques are radio frequency tagging systems, electronic tags, ultrasonic systems \cite{UT} and reflective particle tags \cite{RPT} etc. The main disadvantages of some of the techniques are well analyzed in \cite{Culbreth,D1}, hereafter we only give a short overview of them in the light of the previously defined "intuitive requirements".

Radio frequency systems (RF) consist of a memory chip, an antenna, and a transmitter/receiver system and therefore overcome problems related to printing or etching characters on the side of the container. RF devices can be active or passive. Active tags contain a small internal power source to communicate, store and process large amounts of information in the chip. A power source is usually a lithium battery lasting less than 5 years. This makes them unsuitable for use in long-term storages. Passive tags have no battery. In order to provide power and data to the chip, they use the current in the loop antenna which is induced by the interrogating RF signal. Thus, they receive power from the reader's antenna. However, their main drawback is that only a limited amount of information (roughly a few bits) can be stored. The main problems encountered with both active and passive RF devices is related to interference of the metallization layer with the RF signal, locating methods and low transmission range.

Electronic tag technology contains transponder units that hold a unique identification number retrievable by touching it to an inductively-coupled reader. However, they are able to store up to a few tens of kilobytes of information and work for up to 10 years. When considered for application to long-term stored items (i.e. stored for more than 10 years), e.g. nuclear waste containers or spent fuel copper canisters, present labeling methods listed above fail either in terms of operation time or security, namely risk of falsification or error.

Ultrasonic tagging \cite{UT} is based on the assumption of the uniqueness of the welding area of the cask. Thus, it assumes that in the process of ultrasonic scanning one can obtain a unique fingerprint for each stored container. However, this method is rather young and therefore it is difficult to explicitly evaluate its performance in terms of environmental safety, long operating time and security. The definite drawback of this method is inability to provide a large and unique tag memory.

Reflective particle tags have been proposed by Sandia National Laboratories (SNL) in 1992 \cite{SNL}. The tag represents the transparent adhesive matrix with encapsulated reflective particles. Although this system would be good enough to provide the identification for non-nuclear long-stored waste it will be difficult to apply it to casks containing radioactive material due to the difficulties connected to the reader system (a number of lights which induce the reflection in the tag) and presence of gamma background outside the cask walls. The characteristics of the reflective particle tag regarding long operating times, a large and unique tag memory and security can not be evaluated explicitly due to the present research stage of the technology.

Accordingly there is a recognized need for a labeling system which last at least from ten to a few hundred years (time factor), at the same time enabling fully unique identification of the canister contents in a manner consistent with permanent records of the storage or repository (information factor), have a high level of security, i.e. low risk of falsification or error (security factor), and give the possibility to avoid a corrosion effect of canisters induced by the traditional tagging methods (environmental factor). A potential and drawbacks of a new tungsten-based method was considered in following publications \cite{D2,D3}

\section{The main concept}
The main idea of the proposed method consists of using a unique combination of radioisotopes with different predictable length of life and a long operating time, wherein the unique combination of radioisotopes comprises the mixture of two or more radioisotopes \cite{Patent}.
   \subsection{Enabling a non-contact reader system}

Radioisotopes are atoms with nuclei that decay to a more stable nuclear configuration by emitting radiation. Different radioisotopes emit different types of ionizing radiation, such as gamma (${\gamma}$), neutron (n), alpha (${\alpha}$) and beta (${\beta}$) radiation. The gamma, neutron, alpha and beta radiation have different penetration properties.

For example, alpha particles have very little penetrating power and can be stopped by a sheet of paper; a beta particle is lighter than an alpha particle and can be stopped by a thin sheet of metal. At the same time gamma rays and neutrons can be extremely energetic and highly penetrating, so that several meters of concrete might not be enough to stop them.

Thus, the type of emitting radiation can be selected based on the penetration properties of the radiation and the material of item assigned to the long-term storing or for final repositories and the long-term storage. As an example, for spent fuel waste copper canisters the preference in choice of radioisotopes will be given to radioisotopes emitting gamma rays, which are used frequently in medical applications and in industry to check for cracks or flaws in valves.

Thus, the use of a unique combination of radioisotopes as a tag enable the possibility to have a non intrusive reader system.

   \subsection{Enabling a long operating time}
Another inherent property of radioisotopes is a predictable length of life. The radioisotopes decay over time. The time it takes one-half of the atoms of the radioisotope to decay by emitting radiation is the half-life of the radioisotope. After ten half-lives only one thousandth of the atoms of the radioisotope remains. The half-life of the radioisotopes can range from a few seconds (short-lived radioisotopes, e.g. 54.5 seconds for $^{220}$Rn) to hundreds of years (long-lived radioisotopes, e.g. 432.2 years for $^{241}$Am).

In connection with the half-life of the radioisotope a number of parameters may be determined, such as the number of decays per unit time (total activity), the number of decays per unit time per mass or volume of the radioisotope at a time set to zero (specific activity) and the total number of atoms present. These and other parameters without limitation can be used in order to evaluate the age of radioisotopes in a unique combination of radioisotopes.

Thus, the long operating time, up to a few hundreds years, can be accomplished by the present approach because the method for radioisotope labeling of waste and items assigned to long-term storing or for final repositories includes, but is not limited to, using various radioisotopes with different half-lifes. In regards to the spent fuel copper canister assigned to long-term storing or for final repositories the preference in choice of radioisotopes will be given to long-lived radioisotopes.

   \subsection{Environmental safety (i.e. avoid corrosion effects of copper canisters)}
One operative feature of the present method is that it provides an approach to avoid corrosion effects of canisters induced by traditional tagging methods. Compared to traditional tagging methods, the penetrating properties of radiation emitted by a unique combination of radioisotopes do not require a violation of the integrity of the container material. Thus, it provides a possibility to realize a non-destructive labeling approach to avoid corrosion effects of canisters induced by traditional tagging methods.

\begin{figure}[ht!]
\centering
\includegraphics[width=0.5\textwidth]{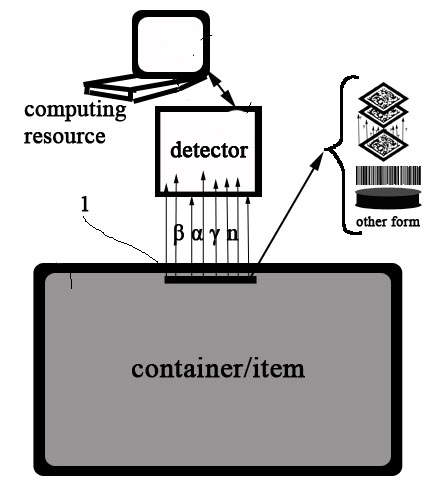}
\caption{An example of how to practically introduce a method of radioisotope labeling.}
\label{fig:3}
\end{figure}

As an example, the embossing of bar codes on the copper layer of the spent fuel copper canister can lead to corrosion of copper in the long-term run and leakage of spent fuel components later on. However, when using the present method, as best seen in Figure \ref{fig:3}, the tag can be placed inside the item, in this specific case, inside the spent fuel copper canister. Thus, a non-destructive labeling approach can be accomplished.
The radiation emitted by the unique combination of radioisotopes, in this specific case, gamma radiation, will penetrate the material of the item, here copper, and can therefore be detected. The type of detector can be chosen based on the type of radiation emitted by the unique combination of radioisotopes. In this specific case, a gamma radiation detector will be chosen. Data obtained by the detector will be provided to the analyzing device to make operational decisions based upon the results. The tag can be arranged in various configurations based on the needs of the identification process. It can be encapsulated in various places inside or outside the item, to optimize use of space and radiation emitted by the unique combination of radioisotopes.

     \subsection{Large and unique tag memory}
Selection of types and quantities of radioisotopes includes determining activity, the age of the radioisotopes, ratios between the lines in the spectra emitted by the radioisotopes in connection with the amount of information to be carried. Among other parameters, the ratios between the lines in the spectrum emitted by radioisotopes can be emphasized. The lines in the spectra emitted by radioisotopes will be one among other unique characteristics of radioisotopes which is used by the method for radioisotope labeling of waste. Each radioisotope has a unique structure of the spectrum emitted (e.g. gamma spectrum). Thus, the lines in the spectra emitted by radioisotopes might be used to identify the presence of a particular radioisotope in the tag. Accordingly, the ratios between the lines of the same or different radioisotopes present in the unique combination of radioisotopes might be used as a unique identifier capable of carring the information about the item assigned to long-term storing or for final repositories.

As a simplified example on how to use the method in order to separate between different types of waste and to determine when the waste was encapsulated or placed in the storage, different radioisotopes may be used in a unique combination of radioisotopes as an identifier of the specific type of waste. For example, the unique combination of radioisotopes of $^{241}$Am (a) + $^{137}$Cs (b) + $^{60}$Co (c) + etc. can be used in a following form but not excluding alternative combinations:
\begin{enumerate}
  \item $^{241}$Am (a) + $^{137}$Cs (b=0) + $^{60}$Co (c=0) + etc. as identifier of liquid waste;
  \item $^{241}$Am (a) + $^{137}$Cs (b) + $^{60}$Co (c) + etc. as identifier of medical waste;
  \item $^{241}$Am (a=0) + $^{137}$Cs(b=0) + $^{60}$Co (c) + etc. as identifier of solid waste.
\end{enumerate}
Parameters a, b, c, etc. refer to the activity or quantity of radioisotopes. In this particular example, every item which has the radioisotope $^{137}$Cs in the unique combination of radioisotopes in the tag will be referred to as medical waste. Thus, in order to identify a canister with medical waste among other items, the combination with radioisotope $^{137}$Cs, i.e.  $^{241}$Am (a) + $^{137}$Cs (b) + $^{60}$Co (c) + etc. will be used. Time characteristics of this cask, such as time when the item was encapsulated, placed in the storage, moved etc., will be evaluated based on the combination of the parameters of the radioisotopes in the unique tag. Among these parameters, relative strength of lines in the spectra emitted by the tag, the half-life of the radioisotopes, activity of the radioisotopes and other characteristics can be used.

The number of radioisotopes in the tag can be limited to a number of unique combinations available in order to label all items to be stored uniquely. Thus, the number of radioisotopes to be used in the unique combination of radioisotopes will be related to the capacity of the storage or repository. In the general case, the affiliation of waste to storage or repository might be also included and identified in the way similar to described above or in any alternative way.

Alternatively, a large and unique tag memory might be obtained by selecting coding technology for the spraying of radioisotopes. Then, the unique combination of radioisotopes might be imprinted or plated or sprayed or distributed in the form of an one-dimensional or two-dimensional matrix symbology or image. A visualization of the one-dimensional or two-dimensional matrix symbology can be provided via detection of radiation emitted by the unique combination of radioisotopes. The information contained in the obtained image or symbology might be extracted or decoded in a way similar to that which reading systems use for decoding of 1D/2D bar codes.

Conventional one-dimensional (1D) bar coding technology is composed of parallel strips holding tens of characters per inch. High speed scanning systems allow this code to be read even from large distances. However, the amount of data encoded in a 1D bar code is limited. Therefore, recently a bi-directional two-dimensional (2D) matrix bar code symbology was developed, where information is encoded through orthogonal lines containing individual bits of information. Thus, 2D bar codes allows carrying hundred times more information in the same space as the 1D bar code. Examples of 2D symbology can be found in \cite{3}.

Bar codes are generally printed or etched onto the side of each canister or item. Therefore, in the present state of art they can be falsified, can eventually lead to the corrosion of container material and leakage of nuclear material. However, the proposed approach of using a unique combination of radioisotopes as a tag allows to avoid the described drawbacks and use the advantages of bar code methodologies.

The amount of unique tag memory of the method for radioisotope labeling of waste might be increased if the strength, the ratios of the lines in the spectra emitted by radioisotopes are utilized in the coding and further analysis of visualization of the one-dimensional and/or two-dimensional matrix symbology. Thus, the three-dimensional (3D) bar code symbology or multidimensional bar code symbology might be realized.

\begin{figure}[ht!]
\centering
\includegraphics[width=0.7\textwidth]{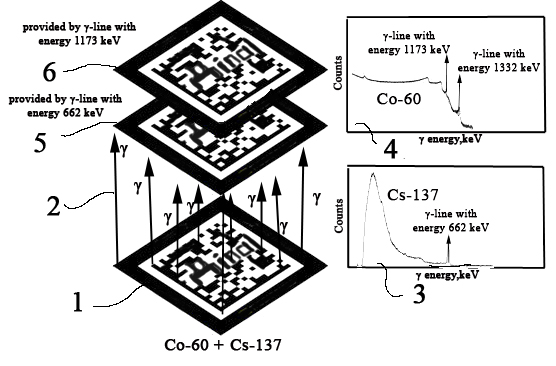}
\caption{An illustration of an example how the three-dimensional (3D) bar code symbology might be achieved with a number of lines in the spectra emitted by isotopes used in a method of radioisotope labeling of waste.}
\label{fig:4}
\end{figure}

The simple illustration of an example how the three-dimensional (3D) bar code symbology might be achieved with a number of lines in the spectra emitted by radioisotopes is best seen in Figure \ref{fig:4}.

The tag 1 in Figure \ref{fig:4} includes the combination of two radioisotopes, imprinted in a form of a two-dimensional matrix symbology. In this particular example, gamma emitting radioisotopes, namely $^{137}$Cs and $^{60}$Co are used. As identifiers gamma radiation 2 is used. The $^{137}$Cs radioisotope can be characterized by the ${\gamma}$-line with energy 662 keV as best seen in the spectrum 3, whereas $^{60}$Co can be characterized by two ${\gamma}$-lines with energy 1173 keV and 1332 keV, as best seen in spectrum 4. Thus, at least two two-dimensional matrix symbology can be realized, 5 provided by the ${\gamma}$-line with energy 662 keV and 6 provided by the ${\gamma}$-line with energy 1173 keV. Each of them can have different or similar information encoded. In a similar way, by using larger amount of radioisotopes and the lines in the spectra emitted by radioisotopes, a multidimensional bar code symbology might be accomplished. As alternative, the unique combination of radioisotopes might be imprinted, plated, sprayed or distributed in the form of image or text.

   \subsection{Security technique against falsification of data, errors/multiple verification}
\begin{figure}[ht!]
\centering
\includegraphics[width=0.7\textwidth]{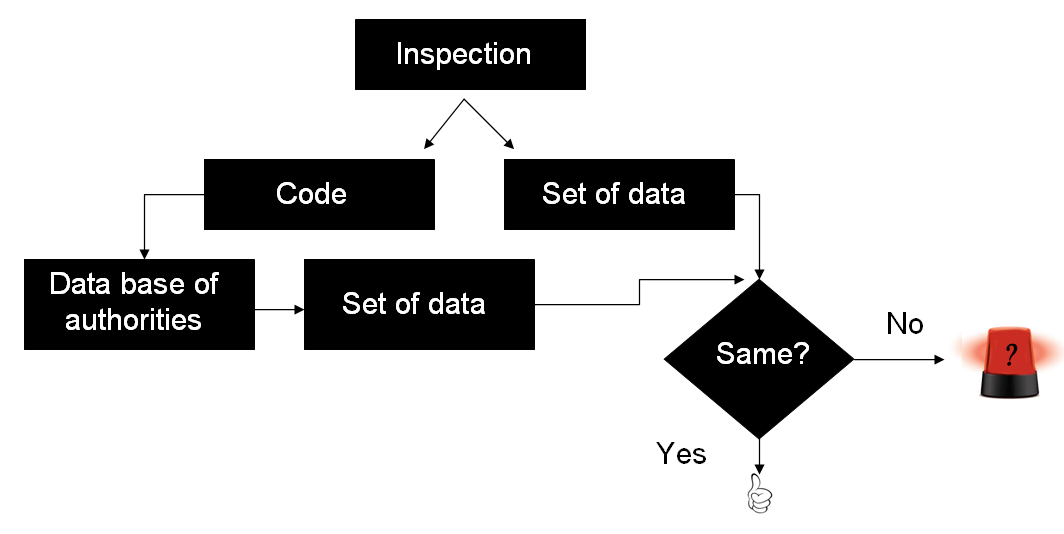}
\caption{A simplified illustration of the security concept against falsification of the database or the tag.}
\label{fig:5}
\end{figure}

Security against falsification of data or errors can be accomplished by using a unique digital code and a list of independently evaluated parameters as shown in Figure \ref{fig:5}.
The unique digital code includes, but is not limited to, determination of a limited sequence of symbols which lend itself to an automated inventory/database system. The upper limit for the sequence of symbols can be chosen based on the capacity of the storage or repository. Further, the unique digital code can be linked to the inventory/database system. Thus, the item in question can be identified. The information which is contained in the inventory/database system may include, but is not limited to the information about the content of the item and information about the item itself. Thus, this information obtained from the inventory/database system may be compared to the list of independently evaluated parameters.

The list of independently evaluated parameters includes determination of parameters describing the item itself and the content of the item. As a simple example, analysis of visualization of, at least, the one-dimensional or multi-dimensional matrix symbology can be used to obtain independently, i.e. not from the inventory/database system, parameters in question. Thus, the timely detection of errors, falsification of the data in the inventory/database system may be provided by comparison of two data sets obtained from independent evaluation (the list of independently evaluated parameters) and from database record through the unique digital code.

  \subsection{Previous use of radioisotopes as tracers}
It is interesting to notice that the idea to use radioisotopes as tracers has been shown to be a robust method in labeling nucleic acids, where radioisotopes are incorporated into the DNA through enzyme action \cite{DNA}. Some studies from the early 60-s also give the example of tagging small animals, e.g. voles, moles, mice and bats etc. with radioisotopes, such as $^{45}$Ca, $^{182}$Ta, $^{124}$Sb, $^{131}$I, $^{198}$Au, $^{60}$Co for tracing purposes \cite{animals}. Later on, radioisotope tracers based on $^{110m}$Ag were applied to the problem of identification of stolen electrical copper cable \cite{copper1}. The technique was demonstrated, but did not get a wide application due to the radiation safety requirements.

\section{Realization of a multidimensional bar code symbology by a unique radioisotope tag}

\begin{figure}[ht!]
\centering
\includegraphics[width=0.70\textwidth]{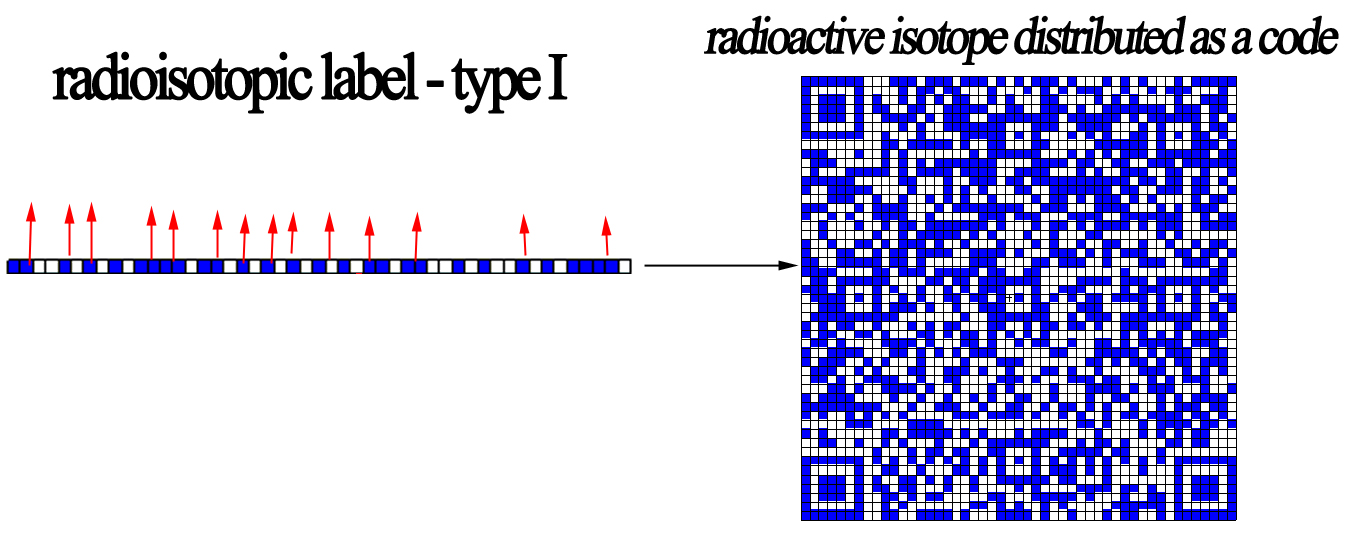}
\caption{A simplified illustration of the implementation of bar code symbology \cite{barcode} by a unique radioisotope tag - type I.}
\label{fig:6}
\end{figure}

As was mentioned above, one option for a new labeling method can be realized through implementation of a multidimensional bar code symbology with a unique combination of radioisotopes or just one single isotope with specific characteristics. Despite the attractiveness of this concept, there are a number of difficulties associated to it. They are related to the handling of the isotopes and providing security against falsification of the tag. In particular, if a unique combination of radioisotopes should be imprinted, plated, sprayed or distributed in the form of a one-dimensional or two-dimensional matrix symbology or image, as shown in Figure \ref{fig:6}, the company/lab/facility must have a permission from the radiation safety authorities and have specific equipment in order to handle radioisotopes. This could lead to time and cost implications.

    \subsection{Pre-defined code and content of information}
However, if the information about the item is known in advance, a series of radioisotope tags can be produced at a facility licensed for this type of work and afterwards used by the encapsulation company.
The operation principle of such a bar code is similar to the original version of the unique radioisotope tag, i.e. radiation emitted by the isotope is detected by using a radiation detector. After that the required parameters (types of isotopes, activity etc.) are estimated. In addition to this, the distribution of radioisotopes in the unique radioisotope tag is measured by using a position sensitive detector or collimator. As a final step, the bar code image is reconstructed and presented.

    \subsection{Non-defined code and information}
\begin{figure}[ht!]
\centering
\includegraphics[width=0.99\textwidth]{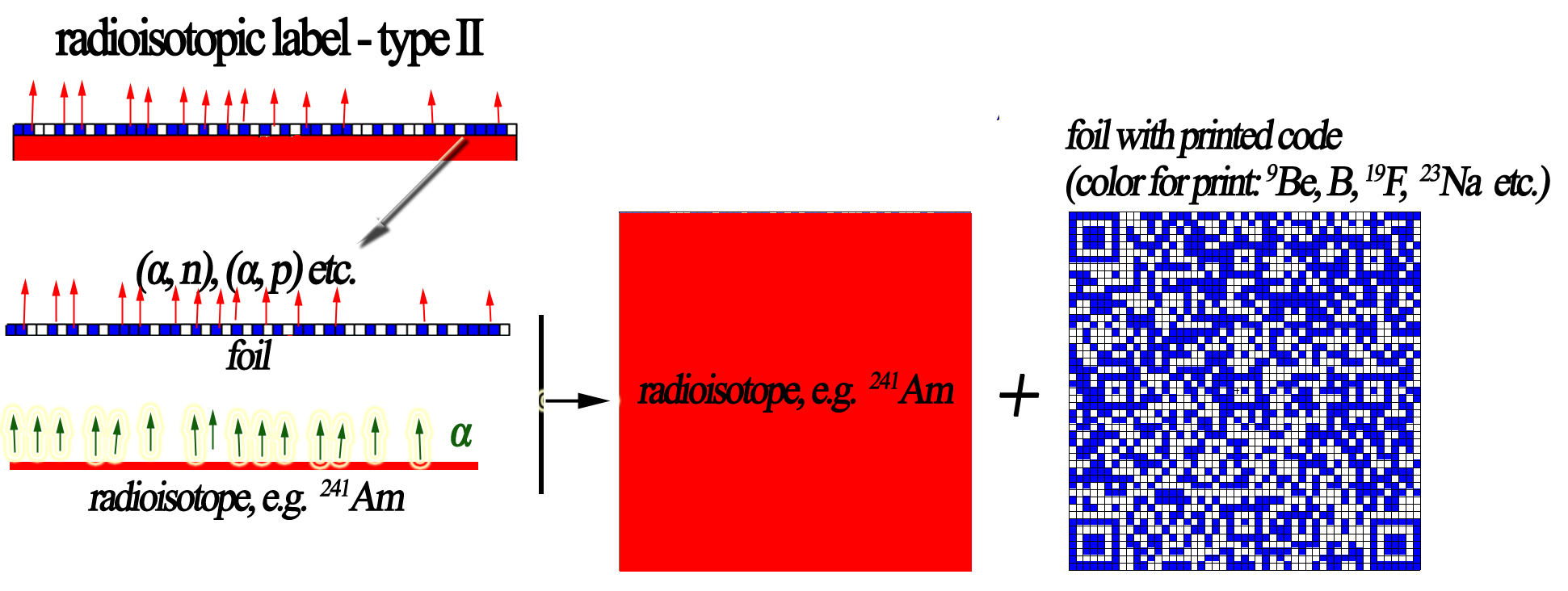}
\caption{A simplified illustration of the concept for providing the multidimensional bar code symbology \cite{barcode} by a unique radioisotope tag - type II.}
\label{fig:7}
\end{figure}

In a situation where the information about the item is not known in advance and should be encoded in the unique radioisotope tag directly at the encapsulation plant, we propose to use another version of the tag, as shown in Figure \ref{fig:7}.
This realization of the tag includes two components: a radioisotope plate prepared by authorities at a facility licensed for this work and a foil which is printed at the encapsulation plant. This concept is appealing in that while printing the tag a specific color could be used. As an example, the base of the tag can be made of an $^{241}$Am or $^{210}$Po ${\alpha}$-emitting isotope which are widely used in smoke detectors. Then the foil which contains the bar code could be printed with colors based on $^{9}$Be, $^{23}$Na, $^{19}$F, $^{10, 11}$B, $^{30}$P, $^{7, 6}$Li etc. materials. After that, the printed foil must be placed in close contact with the ${\alpha}$-emitting base of the tag. These materials have a high cross-section for ${\alpha}$-induced reactions, such as (${\alpha}$,n), (${\alpha}$,p) etc. Thus, the bar code might be read detecting ${\alpha}$-induced gamma rays. The energy of the gamma rays depends on the material which is chosen for printing the tag. As an example, Table 1 shows the energy of the gamma rays which are released in ${\alpha}$-induced reactions on various materials as illustrated in Figure \ref{fig:7}.
\begin{table}[ht]
\caption{Main ${\gamma}$-lines originating from ${\alpha}$-induced reactions in $^{9}$Be, $^{23}$Na, $^{19}$F, $^{10, 11}$B, $^{7}$Li} 
\centering 
\begin{tabular}{c c c c c} 
\hline\hline 
Materials & Energy of ${\gamma}$-rays, keV & Reactions & References  \\ [0.5ex] 
\hline
\emph{$^{7}$Li}     & 478 & $^{7}$Li(${\alpha}$,${\alpha^{\prime}}$ $\gamma$) & \cite{Chaturvedula}, \cite{Croft1}, \cite{Peerani2}     \\
\emph{$^{9}$Be}     & 4439(8) & $^{9}$Be(${\alpha}$,n $\gamma$)& \cite{Chaturvedula}, \cite{Croft2}, \cite{Peerani2}    \\
\emph{$^{19}$F}     & 1275 & $^{19}$F(${\alpha}$,p $\gamma$) & \cite{Chaturvedula}, \cite{Croft1}    \\
\emph{$^{19}$F}     & 197, 1236, 1349, 1357 & $^{19}$F(${\alpha}$,${\alpha^{\prime}}$ $\gamma$) & \cite{Croft1}    \\
\emph{$^{19}$F}     & 583, 637, 891, 1280, 1369, 1401, 1528, 1555 & $^{19}$F(${\alpha}$,n $\gamma$) & \cite{Croft1}    \\
\emph{$^{19}$F}     & 2081, 3182, 3869 & $^{19}$F(${\alpha}$,p $\gamma$) & \cite{Croft1}    \\
\emph{$^{23}$Na} & 1809 & $^{23}$Na(${\alpha}$,p $\gamma$) & \cite{Chaturvedula}  \\
\emph{$^{11}$B}     & 2313 & $^{11}$B(${\alpha}$,n $\gamma$) & \cite{Croft1}, \cite{Peerani2}   \\
\emph{$^{10}$B}     & 3088, 3684, 3854 & $^{10}$B(${\alpha}$,p $\gamma$) & \cite{Croft1}, \cite{Peerani2} \\

\end{tabular}
\label{table:simulat} 
\end{table}

\section{General example of application of a method for steel containers and copper casks}
Here we present first test studies of the new concept. In particular, we consider two general cases, when the tag is applied to steel and copper containers.
\subsection{Description of the containers}
\begin{figure}[ht!]
\centering
\includegraphics[width=0.99\textwidth]{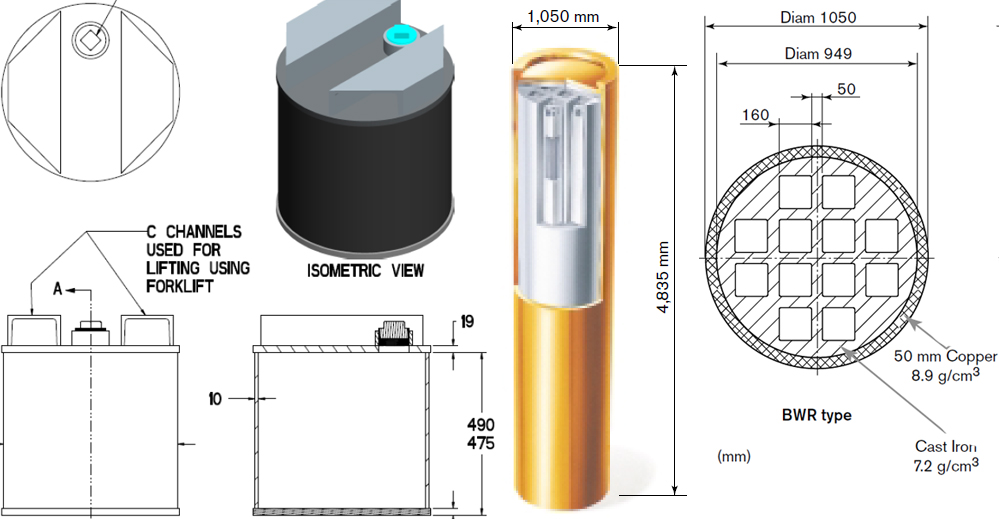}
\caption{Images and dimensions of steel and copper canisters: picture from r.h.s. is taken from \cite{copper} - copper canister with 50 mm wall thickness; picture from l.h.s. is taken from \cite{mercury} - steel canister with 19 mm wall thickness.}
\label{fig:8}
\end{figure}

The steel container chosen corresponds to the one which is commonly used for long-term storing of mercury in Europe and USA \cite{mercury}, while the design of copper canister is the one which is proposed for storing spent fuel in a nuclear repository \cite{copper}, both types are in Figure \ref{fig:8}.

Since normally the tag should be placed inside the container, in the experimental studies we consider two different setups, i.e. a steel block with a thickness of 19 mm and a copper cylinder with a thickness of 10 cm which is twice that of a real cask.  
    \subsection{Description of the experimental studies}
The first test of concept was performed with the use of two types of sources, $^{137}$Cs and $^{60}$Co, as an example to cover the low and high energy regions. In real conditions the choice of isotopes must be done taking into account all details of the industrial application. In the present experiment these sources just played the role as one of the variations of the simple implementation of the radioisotope tag. The gamma signatures were measured by a high purity germanium (HPGe) detector of coaxial type with high voltage bias set to -3000 V.

\begin{figure}[ht!]
\centering
\includegraphics[width=0.99\textwidth]{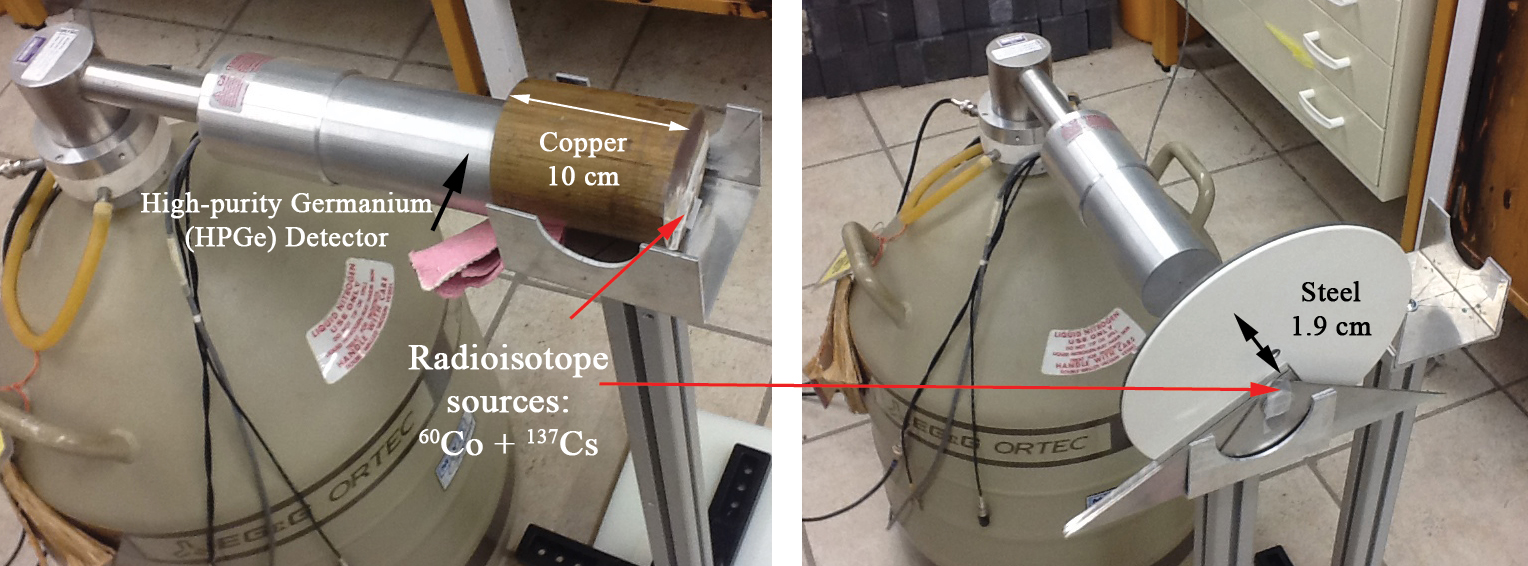}
\caption{Photo and schematic drawing of the configuration of the experimental set-up.}
\label{fig:9}
\end{figure}

The sources were placed approximately 10 cm and 1.9 cm from the detector for the copper and steel container, respectively. The copper cylinder (10 cm in thickness) and steel block (1.9 cm in thickness) were positioned between the detector and sources, as shown in Figure \ref{fig:9}.

\begin{figure}[ht!]
\centering
\includegraphics[width=0.99\textwidth]{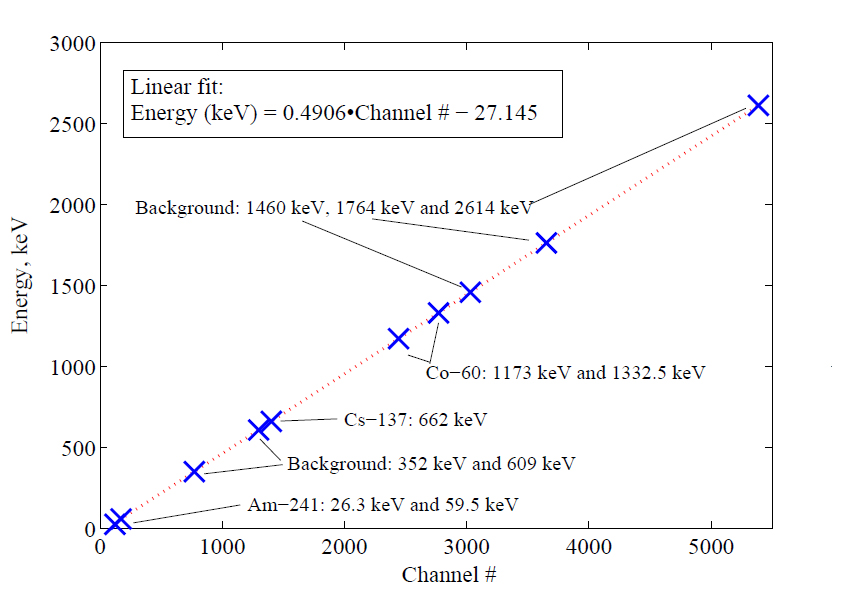}
\caption{The energy calibration of the HPGe detector and linear function fit.}
\label{fig:10}
\end{figure}

\begin{figure}[ht!]
\centering
\includegraphics[width=0.99\textwidth]{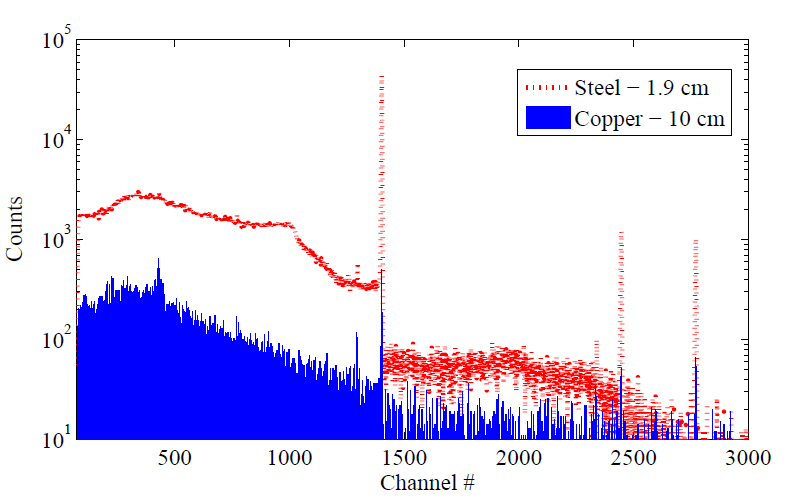}
\caption{Energy distribution of pulses created in the HPGe detector, obtained in the configuration of the experimental set-up.}
\label{fig:11}
\end{figure}

The data was collected during 3646 seconds with a dead time of 1.26 \% for the steel container and during 3617 seconds with a dead time of 0.47 \% for the copper cask. Experimental errors were estimated as in \cite{Boron}. Thus, the live time of measurement was always equal to 3600 seconds. The energy calibration was done by using $^{241}$Am (26.3 keV and 59.5 keV ${\gamma}$-lines), $^{137}$Cs (662 keV ${\gamma}$-line), $^{60}$Co (1173.2 keV and 1332.5 keV ${\gamma}$-lines) calibration standards and background lines, namely 609 keV ($^{214}$Bi), 1460.8 keV ($^{40}$K), 1764 keV ($^{214}$Bi), 2614 keV ($^{208}$Tl). As shown in Figure \ref{fig:10} the linear fit works well at both low and high energies.
The measured spectra with background substraction for both cases (copper and steel) are shown in Figure \ref{fig:11}. The observed ${\gamma}$-lines corresponds to the $^{137}$Cs (662 keV ${\gamma}$-line), $^{60}$Co (1173.2 keV and 1332.5 keV ${\gamma}$-lines) isotopes which made up the tag.
The results are rather straightforward and indicate that the concept of the radioisotope tag can be implemented even with low activity sources. However, it should be mentioned that in the case of application of the radioisotope tag to spent fuel safeguarding, one must be aware of the presence of $^{137}$Cs and $^{60}$Co isotopes in the spent fuel. It should be mention that there are a lot of effort spent in order to evaluate content of the spent fuel prior to encapsulation \cite{System,System2,System3,System4}.
    \subsection{Application of new concept for safeguarding of nuclear waste}
The majority of the background gamma rays in spent fuel originates from activation and fission products, e.g. $^{137}$Cs (662 keV (0.9)\footnote{Here and further branching factor shown in the brackets following the energy.} ${\gamma}$-line), $^{134}$Cs (569 keV (0.15), 605 keV (0.98), 796 keV (0.85), 802 keV (0.09), 1039 keV (0.01), 1168 keV (0.02) and 1365 keV (0.03) ${\gamma}$-lines), $^{144}$Pr (697 keV (0.0148), 1489 keV (0.003) and 2185 keV (0.008) ${\gamma}$-lines), $^{154}$Eu (723 keV (0.19), 873 keV (0.12), 996 keV (0.1), 1005 keV (0.17), 1275 keV (0.36) and 1595 keV (0.03) ${\gamma}$-lines) and $^{106}$Ru (512 keV (0.21), 622 keV (0.1), 1051 keV (0.02), 1128 keV (0.004) and 1357 keV (0.006) ${\gamma}$-lines). Thus, for a fuel cooled for a short period of time ($\sim$ less than four years), the high energy gamma lines, e.g. a 2185 keV gamma line from $^{144}$Pr, will be possible to measure. However, when the fuel will be sent to an encapsulation plant after a number of years of cooling, $^{137m}$Ba, the daughter nuclide of $^{137}$Cs, will be the main gamma emitter. Thus, if the radioisotope tag will have high energy signatures, there will be no problem with radiation background coming from the fuel. Moreover, lead filters can be placed between the detector and the cask in order to suppress low energetic gammas coming from spent fuel placed in the cask. Thus, even the HPGe or CDZT detectors can be used as a reader of the tag.

\begin{figure}[ht!]
\centering
\includegraphics[width=0.99\textwidth]{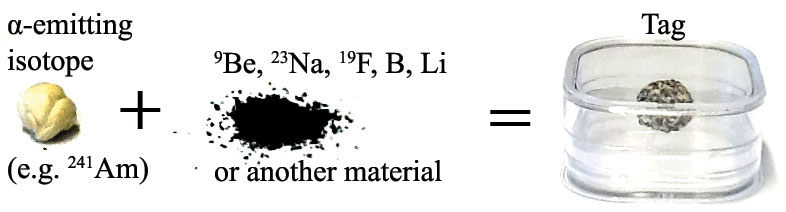}
\caption{The simplest version of the conventional radioisotope tag for safeguarding of nuclear waste.}
\label{fig:12}
\end{figure}

Thus, the simplest version of the conventional radioisotope tag may just include the specific radioisotopes which emits ${\gamma}$-rays with energies higher than 1 MeV. Although, access to these isotopes can be restricted or their cost might be rather high. Therefore, we suggest to use the following version of the radioisotope tag based on the ${\alpha}$-emitting isotopes $^{241}$Am or $^{210}$Po, for example, in a mixture with one of the materials described in Section 3 \emph{"Realization of the multidimensional bar code symbology by the unique radioisotope tag"}. The concept is shown in Figure \ref{fig:12}.
This version of the tag will serve the needs of long-term tagging of nuclear waste, as well as it can solve the existing problem of disposing of smoke detectors or other devices (surge voltage protection devices, electronic valves etc. \cite{EU}) which nowadays contain radioisotopes such as $^{241}$Am. It should be mentioned that according to the Report of the EU commission \cite{EU}, as of the balance sheet date of year 2001, Ireland manufactured 2 million ionization chamber smoke detectors per year (activity of each detector is 33.3 - 37 kBq), while for example Sweden imported 700 000 of them. Thus, the price of the radioisotope tag based on this type of waste will be partly covered by the costs of the waste disposing. At the same time this method will open the possibility of recycling nuclear waste of this type.

\section{First test of the multidimensional bar code symbology}
\begin{figure}[ht!]
\centering
\includegraphics[width=0.99\textwidth]{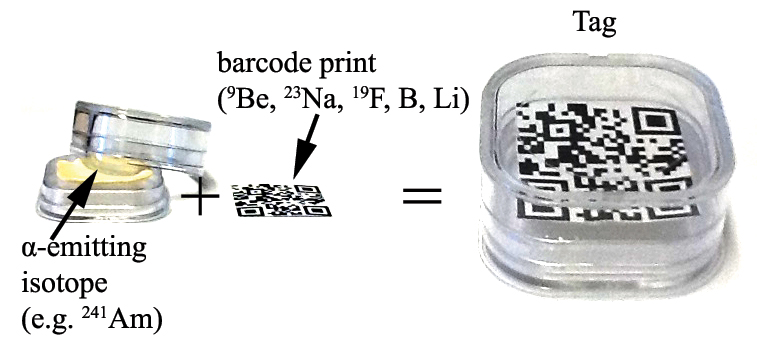}
\caption{A simplified illustration of the radioisotope barcode tag.}
\label{fig:13}
\end{figure}

As we noticed above, one of the attractive options for realization of the radioisotope tag is implementation of the multidimensional bar code symbology, for example in a way as shown in Figure \ref{fig:13}.
In order to explore this alternative the first conceptual studies were done numerically and experimentally, as described below.

Test experiments and simulations were aimed at identifying the possibility to reconstruct the radioisotope bar code. There are two ways how one can do this. The simplest one, which has been used in the present work, is related to the use of a collimator together with a detector. Another method is related to the use of mathematical methods for reconstruction of the image obtained by the detector or position sensitive detector. This requires additional studies and development. However, these two options can be implemented only if there is any observable difference between the results of measurements of the two positions, with source and without it.

    \subsection{Geometry of the simulation and the experimental set-up}
\begin{figure}[ht!]
\centering
\includegraphics[width=0.89\textwidth]{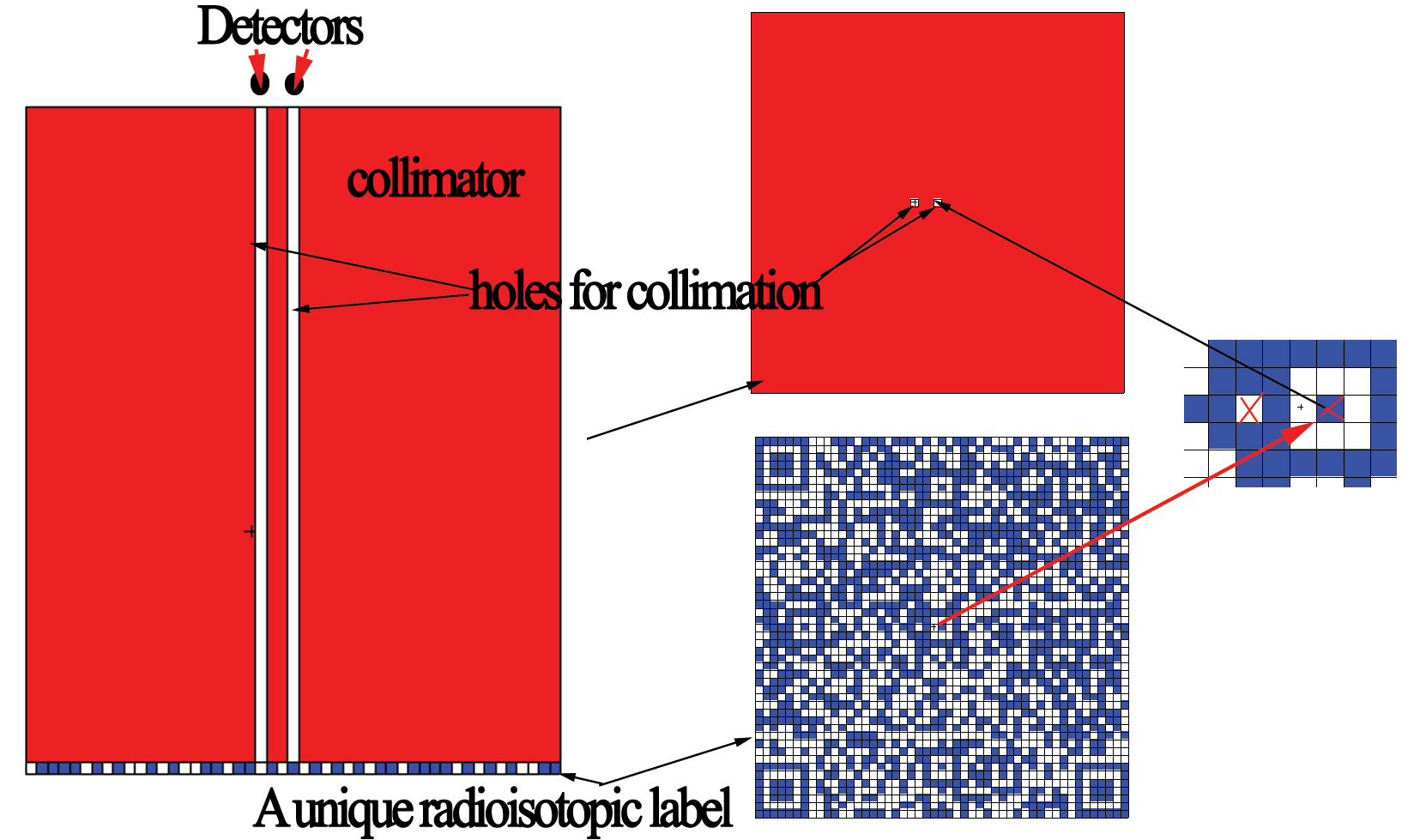}
\caption{A schematic drawing of the configuration of the simulation set-up.}
\label{fig:14}
\end{figure}

Therefore, two extreme cases were considered in both simulations and experiments, i.e. when the measurements position represents the empty space but is surrounded with positions where the source is distributed, and when the position contains the source but is surrounded by empty positions, as shown in Figure \ref{fig:14}.

In order to simulate the realistic situation, information was encrypted in a real barcode \cite{barcode,Carroll}. Afterwards the precise MCNPX \cite{mcnp} model of the barcode was created, as shown in Figure \ref{fig:14}. The blue positions are positions with a distributed source, the white positions are empty positions. One position represents a square with width of 0.1 cm. The total size of the code is 4.9 cm x 4.9 cm. Two point detectors of gamma radiation were placed at a distance of about 5 cm from the top of the tag. Between the tag and the detector, a lead collimator was installed, as shown in Figure \ref{fig:14}. Detector 1 detected the collimated signal from the position which contained the source but was surrounded by the empty positions, detector 2 recorded the signal coming from the empty position.

\begin{figure}[ht!]
\centering
\includegraphics[width=0.45\textwidth]{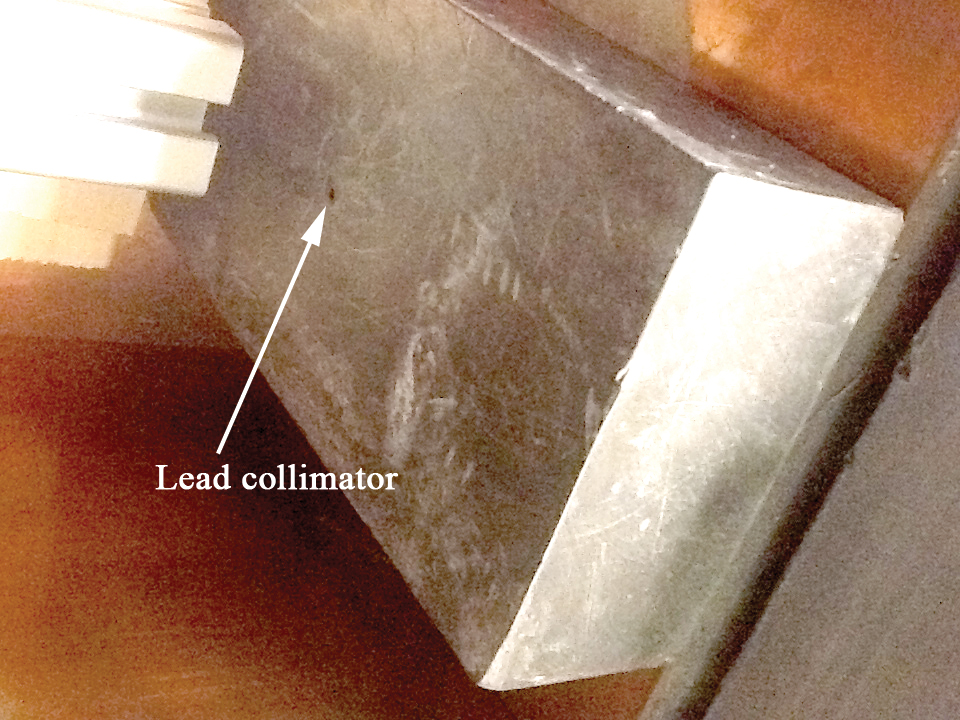}
\caption{A photo of lead collimator with 3 mm hole in the center used in the experiments.}
\label{fig:15}
\end{figure}

The experimental study was organized in a different way, i.e. for a case with a combination of $^{137}$Cs and $^{60}$Co source in the area of the collimation window and for a case where the sources were surrounding the collimation window. The collimator was represented by a lead block with a 3 mm hole in the center, as shown in Figure \ref{fig:15}.

\begin{figure}[ht!]
\centering
\includegraphics[width=0.99\textwidth]{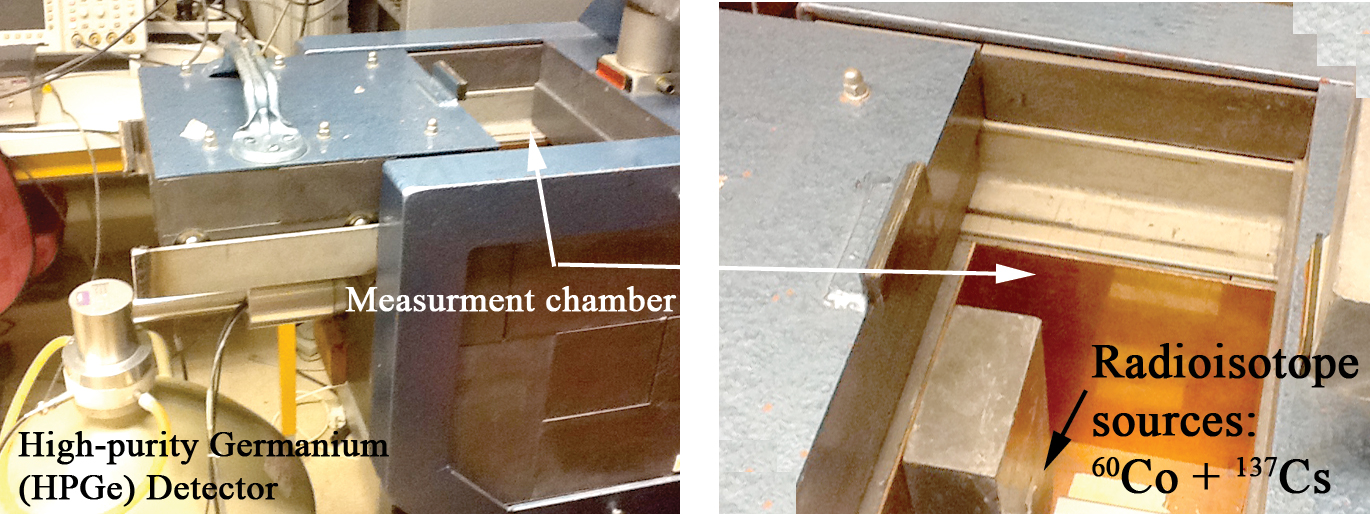}
\caption{A photo of the experimental set-up.}
\label{fig:16}
\end{figure}

In the experiments the HPGe detector was used in the same way and with the same settings as described in the Section 4 \emph{"General example of application of a method for steel containers and copper casks"}. The configuration set-up is shown in Figure \ref{fig:16}.

The data was collected during 12615 seconds with a dead time of 0.12 \% for the case when the sources were in the area of the collimation window and during 12605 seconds with a dead time of 0.04 \% for for the case when the sources were surrounding the collimation window, thus, the live time of the measurement was always equal to 12600 seconds. The energy calibration was done by using a $^{241}$Am (26.3 keV and 59.5 keV ${\gamma}$-lines), $^{137}$Cs (662 keV ${\gamma}$-line), $^{60}$Co (1173.2 keV and 1332.5 keV ${\gamma}$-lines) calibration standards and background lines, namely 609 keV ($^{214}$Bi), 1460.8 keV ($^{40}$K), 1764 keV ($^{214}$Bi), 2614 keV ($^{208}$Tl).

   \subsection{Results of the test and evaluation}
\begin{figure}[ht!]
\centering
\includegraphics[width=0.99\textwidth]{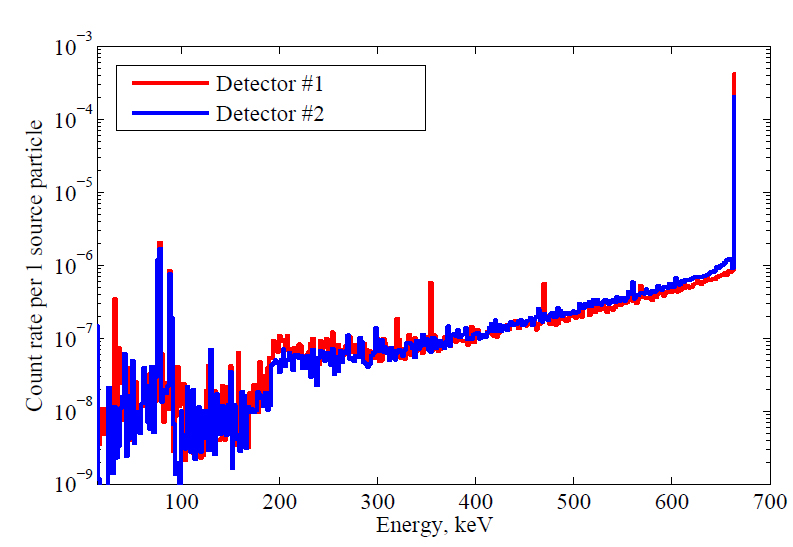}
\caption{Energy distribution of gamma detections obtained in the configuration of the simulation set-up.}
\label{fig:17}
\end{figure}

The simulated gamma spectra for the source (detector 1) and empty (detector 2) positions are shown in Figure \ref{fig:17}.
The line of $^{137}$Cs, which has been used as a source material, is detected in both cases. However, the intensity of the photopeak for the case of measuring the empty space surrounded with the sources is lower than for the case of the source position surrounded by empty positions. This difference is also observed in the structure of the spectra between two measurements in the middle and low energy regions. For example in the low energy region the 34 keV gamma-line is observed only for a case when the collimator is facing the source position.

\begin{figure}[ht!]
\centering
\includegraphics[width=0.99\textwidth]{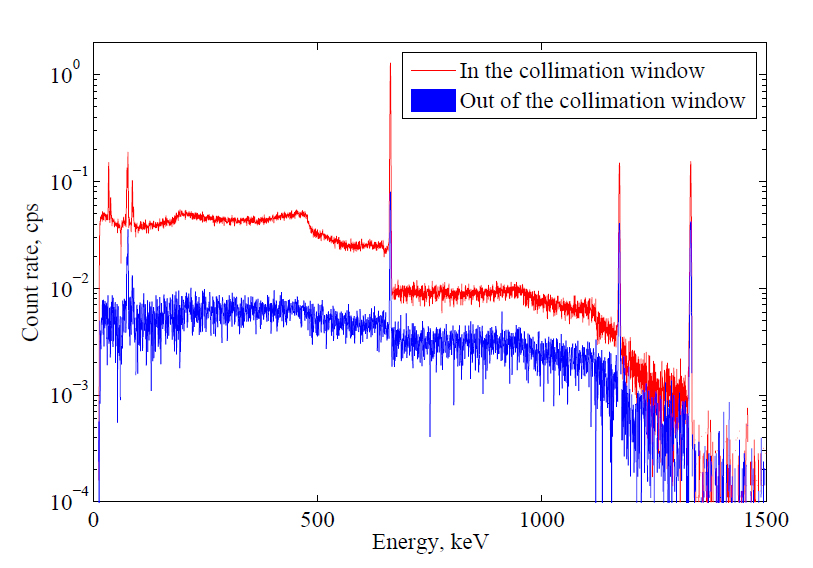}
\caption{Energy distribution of pulses created in a HPGe detector, obtained in the configuration of the experimental set-up.}
\label{fig:18}
\end{figure}

Experimental spectra, shown in Figure \ref{fig:18}, are following the same trend as the simulations. However, they have a better resolution in terms of intensity of the spectra and the photopeaks. This can be due to the width of the collimator. The diameter of the collimation hole was 3 times larger in diameter (3 mm) compared to what was used in the simulations (1 mm). Therefore, further optimization studies are planned to be done in the future work in order to optimize the parameters of the barcode.
It is interesting to notice that the 34 keV gamma-line is observed only for a case when the collimator is facing the source position of the experiment in exactly the same way as it was in the simulations. Thus, we can conclude that there are differences observed in measuring the two extreme cases in both simulations and experiments. This indicates that these differences or specific signatures might be further used in order to reconstruct the barcode.

\section{Conclusions}

In this paper we have described a new concept of long-term security identification tags/labels that can be applied to articles for carrying information about the content, inventory tracking, prevention of falsification and theft etc. The suggested concept is based on the use of unique combinations of radioisotopes with different predictable half life, and the like as a label of the items.

The idea of a new tag was tested with two general designs of storage canisters, namely a steel container which corresponds to the one which is commonly used for long-term storing of mercury in Europe and USA and a copper canister which is the one which is in applications for nuclear repositories. The results of the first test experiments indicate that the concept of the radioisotope tag can be implemented even with low activity sources.

However, in the case of application of the radioisotope tag to spent fuel safeguarding it is suggested to use a mixture of ${\alpha}$-emitting isotopes, such as $^{241}$Am or $^{210}$Po, with materials that easily undergo ${\alpha}$-induced reactions with emission of specific ${\gamma}$-lines, e.g. $^{9}$Be, $^{23}$Na, $^{19}$F, $^{10, 11}$B, $^{30}$P, $^{7, 6}$Li etc. Thus, if the radioisotope tag will have a high energy signature, there will be no problem with radiation background coming from the fuel. Moreover, this version of the radioisotope tag allows to solve the existing problem of the disposing of smoke detectors or other devices \cite{EU} which contain radioisotopes, such as $^{241}$Am, thus, indirectly providing a recycling of nuclear waste. As an economical advantage, it should be mentioned that the price of the radioisotope tag based on this type of waste will be partly covered by the costs of the waste disposing.

As an attractive option for a new labeling method we proposed the possibility to realize a multidimensional bar code symbology. Two different ways of realization are suggested. The first option is when the unique combination of radioisotopes is imprinted, plated, sprayed or distributed in the form of a one-dimensional or two-dimensional matrix symbology or image. Another realization of the tag includes two components: a radioisotope plate prepared by authorities at a facility licensed for this work and a foil which is printed at the encapsulation plant with specific colors based on $^{9}$Be, $^{23}$Na, $^{19}$F, $^{10, 11}$B, $^{30}$P, $^{7, 6}$Li etc. materials. The results of the experiments showed the presence of specific signatures in the spectra which give the possibility to realize a multidimensional radioisotope bar code symbology.

Thus, the new radioisotope label offers several advantages, as compared to the currently used tagging methods. It provides
\begin{enumerate}
  \item Environmental safety.
  \item Non-contact reader system.
  \item Long operating time.
  \item Large and unique tag memory.
  \item Security technique against falsification of data, errors/multiple verification.
  \item Recycling option for ionization chamber smoke detectors and other devices \cite{EU}.
  \end{enumerate}

A further experimental, simulation and mathematical study is planed to be done to optimize and evaluate the different options of the radioisotope label.

\section*{Acknowledgement}
The authors want to thank ASTOR (the Application of Safeguards TO geological Repositories) experts group for useful discussions.

\end{document}